\documentclass[letterpaper,twocolumn,prl,aps,superscriptaddress,amsmath,amssymb,floatfix]{revtex4-1}
\usepackage{mathptmx}
\usepackage[utf8]{inputenc}
\usepackage{color}
\usepackage{amsmath}
\usepackage{amssymb}
\usepackage{graphicx}
\usepackage{esint}
\usepackage{ulem}
\usepackage{siunitx}
\usepackage[T1]{fontenc}
\usepackage[unicode=true,
 bookmarks=true,bookmarksnumbered=false,bookmarksopen=false,
 breaklinks=false,pdfborder={0 0 1},backref=false,colorlinks=true]
 {hyperref}
\hypersetup{ linkcolor=magenta,urlcolor=blue,citecolor=blue,pdfstartview={FitH},hyperfootnotes=false}

\makeatletter

\DeclareFontEncoding{LGR}{}{}

\ProvideTextCommand{\~}{LGR}[1]{\char126#1}

\newcommand{\lyxmathsym}[1]{\ifmmode\begingroup\def\b@ld{bold}
  \text{\ifx\math@version\b@ld\bfseries\fi#1}\endgroup\else#1\fi}

\usepackage{textcomp}
\usepackage{epstopdf}

\usepackage{amsfonts}

\usepackage{soul}

\pdfpageheight\paperheight
\pdfpagewidth\paperwidth

\@ifundefined{textcolor}{}{%
 \definecolor{BLACK}{gray}{0}
 \definecolor{WHITE}{gray}{1}
 \definecolor{RED}{rgb}{1,0,0}
 \definecolor{GREEN}{rgb}{0,1,0}
 \definecolor{BLUE}{rgb}{0,0,1}
 \definecolor{CYAN}{cmyk}{1,0,0,0}
 \definecolor{MAGENTA}{cmyk}{0,1,0,0}
 \definecolor{YELLOW}{cmyk}{0,0,1,0}
}

\usepackage{xcolor}\usepackage{soul}
\definecolor{blue}{rgb}{0,0,1}
\definecolor{red}{rgb}{1,0,0}
\definecolor{green}{rgb}{0,1,0}

\makeatother

\begin{document}

\title{Volcano architecture for scalable quantum processor units}
\author{Dong-Qi~Ma}
\affiliation{Laboratory of Quantum Information, University of Science and Technology of China, Hefei 230026, China}
\affiliation{Anhui Province Key Laboratory of Quantum Network, University of Science and Technology of China, Hefei 230026, China}
\affiliation{CAS Center for Excellence in Quantum Information and Quantum Physics, University of Science and Technology of China, Hefei 230026, China}

\author{Qing-Xuan~Jie}
\affiliation{Laboratory of Quantum Information, University of Science and Technology of China, Hefei 230026, China}
\affiliation{Anhui Province Key Laboratory of Quantum Network, University of Science and Technology of China, Hefei 230026, China}
\affiliation{CAS Center for Excellence in Quantum Information and Quantum Physics, University of Science and Technology of China, Hefei 230026, China}

\author{Ya-Dong~Hu}
\affiliation{Laboratory of Quantum Information, University of Science and Technology of China, Hefei 230026, China}
\affiliation{Anhui Province Key Laboratory of Quantum Network, University of Science and Technology of China, Hefei 230026, China}
\affiliation{CAS Center for Excellence in Quantum Information and Quantum Physics, University of Science and Technology of China, Hefei 230026, China}

\author{Wen-Yi~Zhu}
\affiliation{Laboratory of Quantum Information, University of Science and Technology of China, Hefei 230026, China}
\affiliation{Anhui Province Key Laboratory of Quantum Network, University of Science and Technology of China, Hefei 230026, China}
\affiliation{CAS Center for Excellence in Quantum Information and Quantum Physics, University of Science and Technology of China, Hefei 230026, China}

\author{Yi-Chen~Zhang}
\affiliation{Laboratory of Quantum Information, University of Science and Technology of China, Hefei 230026, China}
\affiliation{Anhui Province Key Laboratory of Quantum Network, University of Science and Technology of China, Hefei 230026, China}
\affiliation{CAS Center for Excellence in Quantum Information and Quantum Physics, University of Science and Technology of China, Hefei 230026, China}

\author{Hong-Jie~Fan}
\affiliation{Laboratory of Quantum Information, University of Science and Technology of China, Hefei 230026, China}
\affiliation{Anhui Province Key Laboratory of Quantum Network, University of Science and Technology of China, Hefei 230026, China}
\affiliation{CAS Center for Excellence in Quantum Information and Quantum Physics, University of Science and Technology of China, Hefei 230026, China}

\author{Xiao-Kang~Zhong}
\affiliation{Laboratory of Quantum Information, University of Science and Technology of China, Hefei 230026, China}
\affiliation{Anhui Province Key Laboratory of Quantum Network, University of Science and Technology of China, Hefei 230026, China}
\affiliation{CAS Center for Excellence in Quantum Information and Quantum Physics, University of Science and Technology of China, Hefei 230026, China}

\author{Guang-Jie~Chen}
\affiliation{Laboratory of Quantum Information, University of Science and Technology of China, Hefei 230026, China}
\affiliation{Anhui Province Key Laboratory of Quantum Network, University of Science and Technology of China, Hefei 230026, China}
\affiliation{CAS Center for Excellence in Quantum Information and Quantum Physics, University of Science and Technology of China, Hefei 230026, China}

\author{Yan-Lei~Zhang}
\affiliation{Laboratory of Quantum Information, University of Science and Technology of China, Hefei 230026, China}
\affiliation{Anhui Province Key Laboratory of Quantum Network, University of Science and Technology of China, Hefei 230026, China}
\affiliation{CAS Center for Excellence in Quantum Information and Quantum Physics, University of Science and Technology of China, Hefei 230026, China}

\author{Tian-Yang~Zhang}
\affiliation{Laboratory of Quantum Information, University of Science and Technology of China, Hefei 230026, China}
\affiliation{Anhui Province Key Laboratory of Quantum Network, University of Science and Technology of China, Hefei 230026, China}
\affiliation{CAS Center for Excellence in Quantum Information and Quantum Physics, University of Science and Technology of China, Hefei 230026, China}

\author{Xi-Feng~Ren}
\affiliation{Laboratory of Quantum Information, University of Science and Technology of China, Hefei 230026, China}
\affiliation{Anhui Province Key Laboratory of Quantum Network, University of Science and Technology of China, Hefei 230026, China}
\affiliation{CAS Center for Excellence in Quantum Information and Quantum Physics, University of Science and Technology of China, Hefei 230026, China}
\affiliation{Hefei National Laboratory, University of Science and Technology of China, Hefei 230088, China}

\author{Liang~Chen}
\email{cliang911@ustc.edu.cn}
\affiliation{Laboratory of Quantum Information, University of Science and Technology of China, Hefei 230026, China}
\affiliation{Anhui Province Key Laboratory of Quantum Network, University of Science and Technology of China, Hefei 230026, China}
\affiliation{CAS Center for Excellence in Quantum Information and Quantum Physics, University of Science and Technology of China, Hefei 230026, China}

\author{Zhu-Bo~Wang}
\email{zbwang@ustc.edu.cn}
\affiliation{Laboratory of Quantum Information, University of Science and Technology of China, Hefei 230026, China}
\affiliation{Anhui Province Key Laboratory of Quantum Network, University of Science and Technology of China, Hefei 230026, China}
\affiliation{CAS Center for Excellence in Quantum Information and Quantum Physics, University of Science and Technology of China, Hefei 230026, China}

\author{Guang-Can~Guo}
\affiliation{Laboratory of Quantum Information, University of Science and Technology of China, Hefei 230026, China}
\affiliation{Anhui Province Key Laboratory of Quantum Network, University of Science and Technology of China, Hefei 230026, China}
\affiliation{CAS Center for Excellence in Quantum Information and Quantum Physics, University of Science and Technology of China, Hefei 230026, China}
\affiliation{Hefei National Laboratory, University of Science and Technology of China, Hefei 230088, China}

\author{Chang-Ling~Zou}
\email{clzou321@ustc.edu.cn}
\affiliation{Laboratory of Quantum Information, University of Science and Technology of China, Hefei 230026, China}
\affiliation{Anhui Province Key Laboratory of Quantum Network, University of Science and Technology of China, Hefei 230026, China}
\affiliation{CAS Center for Excellence in Quantum Information and Quantum Physics, University of Science and Technology of China, Hefei 230026, China}
\affiliation{Hefei National Laboratory, University of Science and Technology of China, Hefei 230088, China}

\date{\today}
\begin{abstract}
Quantum information processing platforms based on array of matter qubits, such as neutral atoms, trapped ions, and quantum dots, face significant challenges in scalable addressing and readout as system sizes increase. Here, we propose the “Volcano” architecture that establishes a new quantum processing unit implementation method based on optical channel mapping on a arbitrarily arranged static qubit array. To support the feasibility of Volcano architecture, we show a proof-of-principle demonstration by employing a photonic chip that leverages custom-designed three-dimensional waveguide structures to transform one-dimensional beam arrays into arbitrary two-dimensional output patterns matching qubit array geometries. We demonstrate parallel and independent control of 49-channel with negligible crosstalk and high uniformity. This architecture addresses the challenges in scaling up quantum processors, including both the classical link for parallel qubit control and the quantum link for efficient photon collection, and holds the potential for interfacing with neutral atom arrays and trapped ion crystals, as well as networking of heterogeneous quantum systems.

\vspace{1em}
\noindent\textbf{Keywords:} Scalable quantum processor, Quantum computing, Photonic chip, Optical channel mapping
\end{abstract}
\maketitle

\section{Introduction}
In the past decades, remarkable progress has been achieved in realizing quantum information science, with recent breakthroughs demonstrating substantial strides in the manipulating and controlling qubits. Among various physical platforms, room-temperature qubit arrays, such as neutral atoms, trapped ions, quantum dots, and color centers, offers a promising route for realizing scalable and controllable quantum processing units (QPUs). These systems benefit from the coherent optical photon-qubit interaction, which enables high-fidelity quantum control~\cite{Levine2019,Evered2023,Harty2014,Clark2021,Press2008,Abobeih2022}, fast readout~\cite{Gibbons2011,Shea2020,Todaro2021,Chow2023,Shields2015,Xu2021,PhysRevLett.134.240802}, and qubit-photon entanglement for further distribution of quantum information~\cite{Beugnon2006,Beukers2024,Stas2022,Knaut2024,Grinkemeyer2025,Main2025,Shi2022,PhysRevLett.130.173601}.
Notable milestones include the successful operation of systems with over 500 trapped ions~\cite{Guo2024} and more than 6,100 neutral atoms~\cite{Manetsch2024}, which have greatly advanced the capabilities of quantum simulation~\cite{Bernien2017,Ebadi2021,Scholl2021,Labuhn2016,Wu2024,Chen2023,Emperauger2025,Jiang2012,Dhawan2024,Du2016,Zhao2025}, machine learning~\cite{Fujii2017,Bravo2022,Martinez-Pena2021}, and optimization algorithms~\cite{Farhi2014,Kandala2017,Kokail2019,Cerezo2021,Zhou2020,PhysRevLett.131.103601}. These advancements have laid a strong foundation for QPUs with large-scale and the further networking of QPUs, making it possible to execute complex algorithms for both near-term quantum applications and long-term fault-tolerant quantum computing.

However, realizing practical, large-scale quantum processors present a series of daunting challenges that must be overcome. A key obstacle is the difficulty of providing precise, local control over individual qubits in large arrays~\cite{Graham2022,Hou2024}, especially as the qubit numbers increase into hundreds or thousands. Achieving efficient, localized addressing of each qubit becomes increasingly difficult in 2D or 3D arrays of qubits, where the spatial distribution of qubits can be irregular~\cite{Guo2024,Holewa2024}. Alongside this, it is also required for fast, non-demolition, high-fidelity readout of individual qubits in practice for mid-circuit measurements and feedforward control~\cite{Fowler2012,Horsman2012,Ma2023,Singh2023,Bluvstein2023}.
Furthermore, another critical challenge lies in achieving the addressed collection of photons that enables the long-distance communication between remote QPUs, a necessary step for distributed quantum computing~\cite{Main2025,Ramette2024,Sinclair2025}.

To address the problem of local qubit addressing, recent advancements have introduced innovative technologies such as Acousto-Optic Deflectors (AOD)~\cite{Olsacher2020,Graham2022,Bluvstein2023,Hou2024}, Digital Micromirror Devices (DMD)~\cite{Zhang2024}, and fiber arrays~\cite{Li2024}, which enable efficient spatial control of laser beams across large qubit arrays. These methods allow for dynamic, addressable optical paths to each qubit, supporting parallel operations and reducing the complexity of qubit manipulation in large systems. More recently, planar photonic chips have been employed for the mapping between individual waveguide output ports and trapped ions~\cite{ionchip,Craft2024,Sotirova2024}, though limited to one-dimensional qubit array. 
Moreover, in terms of photon collection and qubit readout, EMCCD and CMOS cameras are employed for parallel and high-fidelity readout of qubit arrays~\cite{Myerson2008,Bakr2009,Manetsch2024,McMahon2024,Guo2024}. 
In particular, their application in atomic array is limited by the trade-off between the long detection period and the loss of qubits~\cite{Bluvstein2023,Manetsch2024,Li2025, k2w2-83kc}. 
More recently, the moving tweezer to reconfigure the qubit array pattern has been utilized to overcome the addressing control and readout challenges, and the parallel implementation of 48 logical qubits protected by error-correcting codes has been successfully demonstrated~\cite{Bluvstein2023}. Despite these advancements, these technologies remain impeded  by issues such as the speed and accuracy of addressing large numbers of qubits simultaneously and the practical constraints of photon collection in larger systems. 

In this paper, we present an innovative solution to the challenges of scaling qubit arrays in room-temperature quantum processors by proposing the Volcano architecture to support the mapping of large qubit arrays to optical channel arrays. Volcano architecture enables scalable, high-fidelity optical mapping between arbitrarily arranged static qubit arrays and standard optical channels, facilitating precise addressing, efficient qubit control, readout, and interconnection for large-scale quantum processors. Furthermore, the architecture provides an alternative architecture in comparison with other relevant architectures, such as previous demonstrations of atom-based QPU implementation rely on the transportation of atoms in an array, such as the moving tweezer architecture for neutral atom array ~\cite{Bluvstein2023} and the ion shuttling in QCCD architecture for trapped ions ~\cite{Quantinuum2021}. 
In particular, the implementation with photonic chip presented here offers a highly scalable solution with low loss, making it an ideal candidate for high-density qubit array implementations. We propose a photonic chip that incorporates AOD-based addressing and camera-free detection to efficiently handle the real-time readout and long-distance communication of individual qubits within large arrays. This chip achieves a novel combination of scalable qubit addressing and high-fidelity readout, enhancing both local control and interconnectivity among qubits in large quantum processors. By enabling the modular expansion of qubit arrays, our approach also supports the integration of multiple photonic chips, offering a potential solution to scale quantum processors beyond traditional limits. The Volcano architecture that integrates these features represents a significant step forward in the pursuit of scalable, robust, and efficient quantum computing systems, capable of realizing the potential of large-scale quantum applications.

\begin{figure}[t]
\centering
\includegraphics[width=\linewidth]{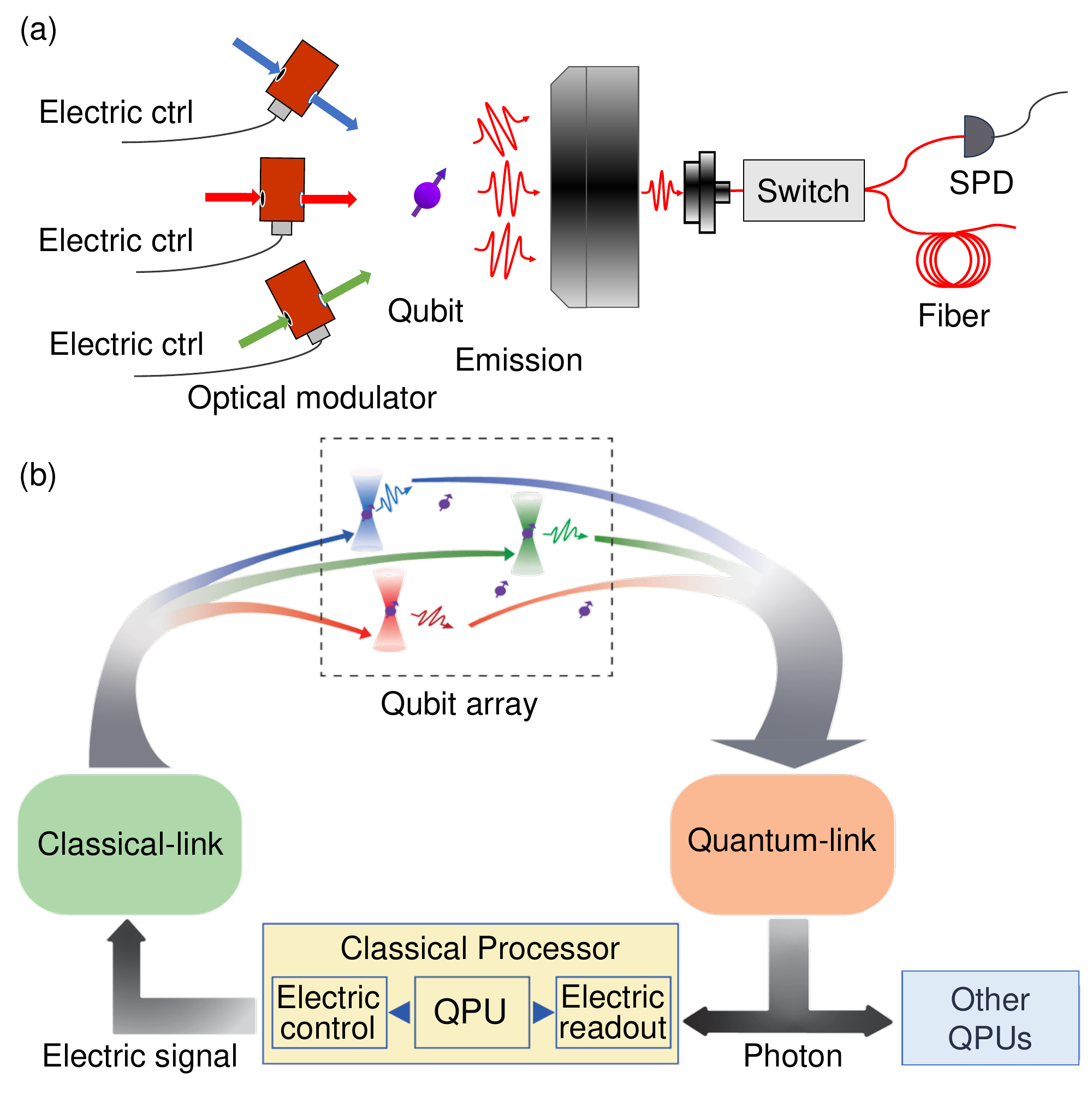}
\caption{\textbf{The element and scale-up of quantum processor units.} 
(a) Schematic representation of the basic setup for single qubit control and readout. Electric signals and continuous wave lasers are used to control the external and internal degrees of freedom of the qubits, and the qubits entangle with single photons through emission, and their quantum states can be detected or entangled with other qubits by collecting and directing the photons into optical fibers. 
(b) Illustration of the scaling challenges in quantum processors, highlighting the need for efficient addressing control and photon collection for individual qubits in an two-dimensional array. The classical link (depicted on the left) provides the necessary optical signals to control and manipulate individual qubits in the quantum processor. The quantum link (depicted on the right) enables the coupling of qubits to photons for measurement and the distribution of entanglement among QPUs. The readout process involves detecting photon emissions from qubits through a fiber array and single-photon detectors.}
\label{fig1}
\end{figure}

\begin{figure*}[t]
\centering
\includegraphics[width=\linewidth]{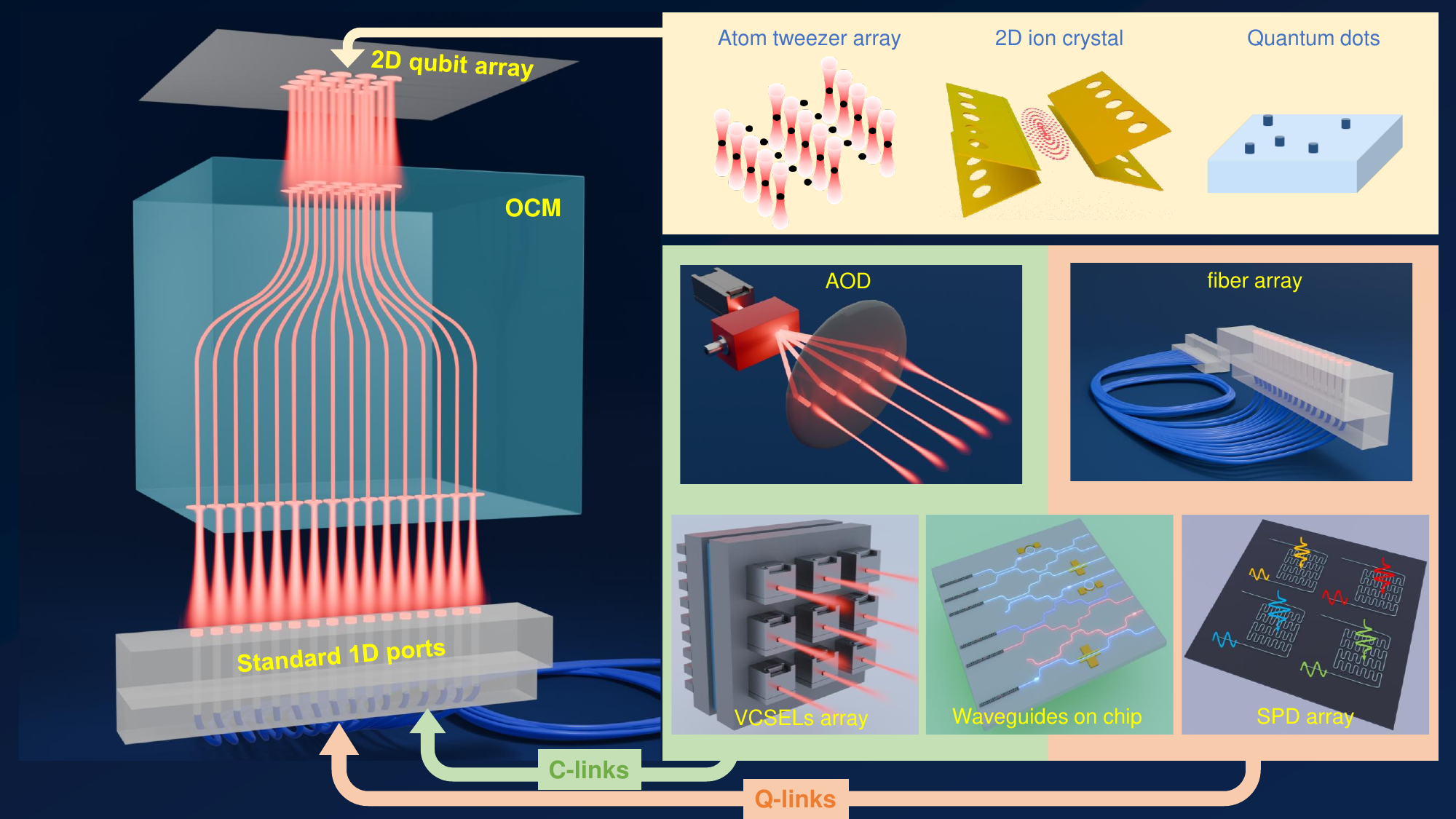}
\caption{\textbf{Volcano architecture.} Left panel: overview of the Volcano architecture, showing how a 2D qubit array is mapped onto a 1D optical channel network via an optical channel mapping (OCM) component. 
Upper-right panel: candidate platforms for realizing the qubit array, including: neutral atoms in optical tweezer arrays; 2D ion crystals; spin qubits implanted in substrates. 
Lower-right panel: available optical elements with massive independent channels for distributing, manipulating, and detecting photons, including:
For C-links: reconfigurable 1D laser beam arrays generated by AODs, vertical cavity surface emitting lasers (VCSEL) arrays; For Q-links:
standard commercial fiber array, single-photon detector (SPD) arrays; For both: planar waveguide arrays on high-refractive index contrast photonic chips. 
 }
\label{fig2}
\end{figure*} 

\section{Volcano Architecture}

Figure~\ref{fig1} illustrates the basic structure of a QPU for room-temperature qubits. 
In general, optical photons provide an indispensable approach to writing and reading the quantum information from qubits, as shown in Fig.~\ref{fig1}(a). Photons serve as information carriers because they can interact readily with qubits, can travel long distances with low loss, and are amenable to being precisely manipulated via various optical techniques. 
On the one hand, electromagnetic waves at optical frequencies enable fast radio-frequency (RF) modulation of carriers, free-space long-distance low-loss signal delivery, and micrometer-scale resolution for addressing individual qubits without the limitation of wiring. Thus, optical fields are often modulated electrically to control various photonic degrees of freedom, enabling initialization and manipulation of the qubit's internal electronic states. Besides, optical photons can also couple to the motional degrees of freedom of qubits, facilitating atom trapping and cooling, making the qubit system stable. 

On the other hand, a coherent qubit-photon interface enables entanglement between qubit states and the photons, which is essential for both quantum state readout and long-range quantum communication. By collecting photons emitted by individual qubits, state measurements can be performed via the number of photons over a certain time. Besides, the entanglement can be distributed over long distances for quantum network applications or used to scale-up the QPUs through distributed quantum computation. 

Figure~\ref{fig1}(b) describes a technical abstraction for scaling QPUs, detailing the core principles behind the individual control and readout of a qubit array. 
The framework consists of four main components: the qubit array, the classical processors, and two crucial channels, i. e. the classical-link (C-link) and the quantum-link (Q-link). 
The classical-link can be treated as a data bus cable bundle, which is responsible for the precise addressing and control of each individual qubit within the quantum processor. It converts conventional electrical signals from the classical processor (e.g., CPUs or FPGAs) into optical carriers. As the number of qubits in the system grows, classical control faces an inherent challenge: the number of control lines scales linearly with qubits, leading to wiring complexity and increased cross-talk. Moreover, scalable QPU development places ever-greater demands on the C-link to curb circuit complexity, for example, the ability to address and renew the control pulse with a short latency on the order of gate durations and to apply parallel control to arbitrary subsets of qubits. 

The quantum-link serves as an adaptor that provides the QPU's quantum input and output channels, interfacing the qubit array with optical networks to transmit quantum information to external devices. Photons then propagate through the system, where they can be routed to remote QPUs or detected and processed by local classical processors. 
Since QPUs demand a finite interaction distance to implement multi-qubit gates, the distance between adjacent qubits in a large-scale QPU is typically only a few micrometers.  In contrast, available optical components that offer on the order of 100 independent channels have restricted spatial arrangement, with fixed pixel sizes, distances, and patterns. 
Therefore, both C-links and Q-links face the challenge of mapping multi-channel optical inputs and outputs to and from the target qubit array pattern, ensuring each qubit corresponds to a distinct optical channel. 
Additionally, as the qubits expand into a 2D or 3D array, accurately addressing each qubit without crosstalk from neighboring qubits or optical channels becomes increasingly challenging. Thus, scaling a QPU is not simply a matter of adding more qubits; it requires simultaneously expanding both the classical and quantum links to support the larger system while preserving operational integrity. These links must grow in tandem to ensure quantum state transfer mechanisms remain robust, low-loss, and high-fidelity as the processor’s size increases.

The solution lies in developing advanced mapping systems that convert a QPU’s 2D qubit array into a matching optical network, enabling the precise control, rapid detection, and long-range communication required for large-scale quantum computing. This is exactly what the Volcano architecture accomplishes via its optical channel mapping (OCM) component. 
Figure~\ref{fig2} illustrates the concept of the Volcano architecture, and the left panel provides a detailed schematic of how quantum information is mapped from a 2D qubit array onto standard optical channels (1D optical fiber array). 
For efficient state control and photon collection, it is preferable to use a one-dimensional beam array that matches the individual channels of a standard AOD and fiber array. For instance, $10\times10$ atom array forms a square lattice of 
$\qty{50}{\mu m}\times\qty{50}{\mu m}$, and the 100-port standard fiber array has a width of $12.7\,\mathrm{mm}$, thus the 3D optical beam mapper provides an array of focusing pipelines for guiding the optical light, resembles a ``Volcano".  This architecture integrates both the quantum and optical subsystems in a modular and scalable manner, allowing for the translation of complex quantum operations onto optical components, which are pivotal for achieving large-scale quantum computing. 

As shown in the upper-right panel of Fig.~\ref{fig2}, qubit arrays have been realized on different physical platforms, such as neutral atoms, ions, or quantum dots, which are typically arranged in two dimensions to build blocks of the quantum processor. These qubits carry quantum information and need to be individually addressed and measured to perform quantum operations. 
The OCM, a central element of this architecture, provides a direct one-to-one bridge between the qubit array and the optical network. OCM comprises an array of low-crosstalk channels, each dedicated to a single qubit and its corresponding optical port.  The channels may take the optical devices of fiber arrays, addressing lasers, or one-dimensional beam arrays generated by AODs. Consequently, OCM serves both the C-link and the Q-link.

In the lower-right panel of Fig.~\ref{fig2}, several commercially available optical elements are shown that provide large‐scale arrays with independent channel management. For example, AOD, vertical cavity surface emitting lasers (VCSEL) array, and waveguides on photonic chip are candidate solutions for implementing C-links by providing configurable laser beam arrays. Meanwhile, fiber array, single photon detector (SPD) array, and photonic chip serve as potential components for Q-links by enabling photon transmission and processing. 
These devices share several characteristics, including compact arrangement, scalability (handle large numbers of independent optical channels, typically on the order of tens to hundreds of channels), parallelism, and standardized patterns (commercially available with standardized geometric or patterns). Whether based on diffraction, interference, waveguiding, or acoustic manipulation, these devices are designed to break down light into individually addressable channels for control or routing. These commercially available elements typically feature standard array patterns, including linear, two-dimensional, or even more complex configurations, each enabling simultaneous photon processing or measurement across many paths. When integrated with the OCM, these elements facilitate efficient photon routing, precise spatial control, and scalable parallel operations by handling multiple optical channels simultaneously. 
Furthermore, combining the OCM with these commercially available optical elements holds significant potential for advanced integration into compact, chip-based systems. Multiple modules can be integrated into the Volcano architecture by connecting the standard fiber arrays and the OCM, further enhancing their practicality and scalability for real-world applications. 

This architecture provides a modular, scalable solution for large-scale quantum processors. As quantum processor size grows, the need for modular optical components that can be interconnected easily becomes crucial. 
The mapping component supports this by enabling the connection of multiple photonic chips into a larger, unified quantum processor. This modular approach is essential for addressing the scalability challenges of quantum computers, as it allows quantum processors to be built in a hierarchical fashion, where smaller, manageable modules can be connected to form larger systems. Each module is responsible for a portion of the total quantum operations, such as quantum entanglement, superdense coding, and quantum error correction, while the central control system coordinates interactions between modules. This modular design prevents a single point of failure, allowing for redundancy and fault tolerance through the simple replacement of standardized optical elements.

\begin{figure*}[!t]
\centering
\includegraphics[width=\linewidth]{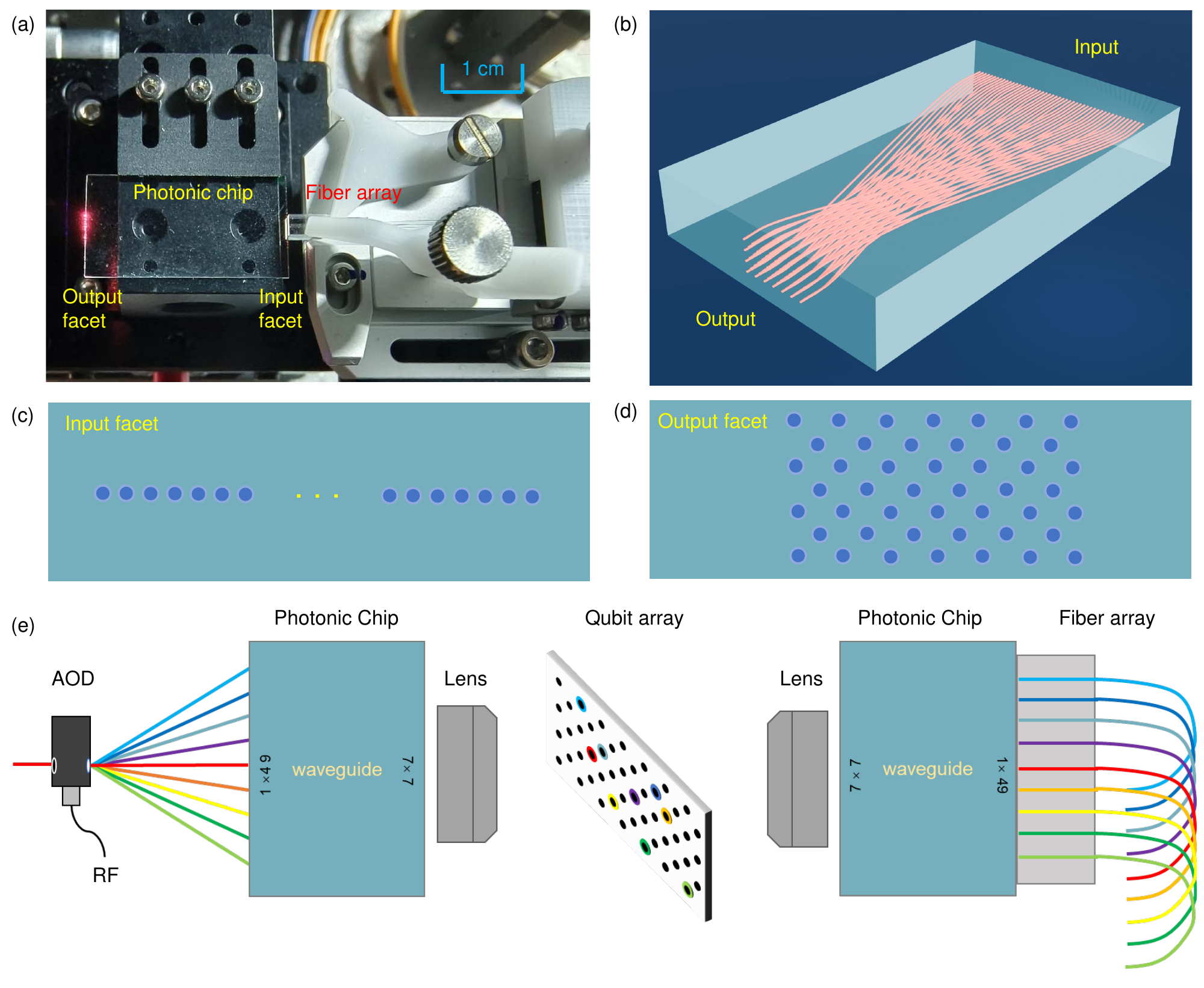}
\caption{3D photonic chip for Volcano architecture implementation. (a) A 3D photonic chip fabricated using laser direct writing technology and placed on motorized stages (on the left), with coupling between the photonic chip and a fiber array (on the right). (b) Topology design of 49 waveguides in the photonic chip. (c-d) Schematics of the input and output facets of the 3D photonic chip. (e) One implementation method of the Volcano architecture is based on the 3D photonic chip. Different colors represent distinct channels for different qubits. The left shows classic optical channels, while the right shows quantum optical channels. }
\label{fig3}
\end{figure*} 

\begin{figure*}[!t]
\centering
\includegraphics[width=\linewidth]{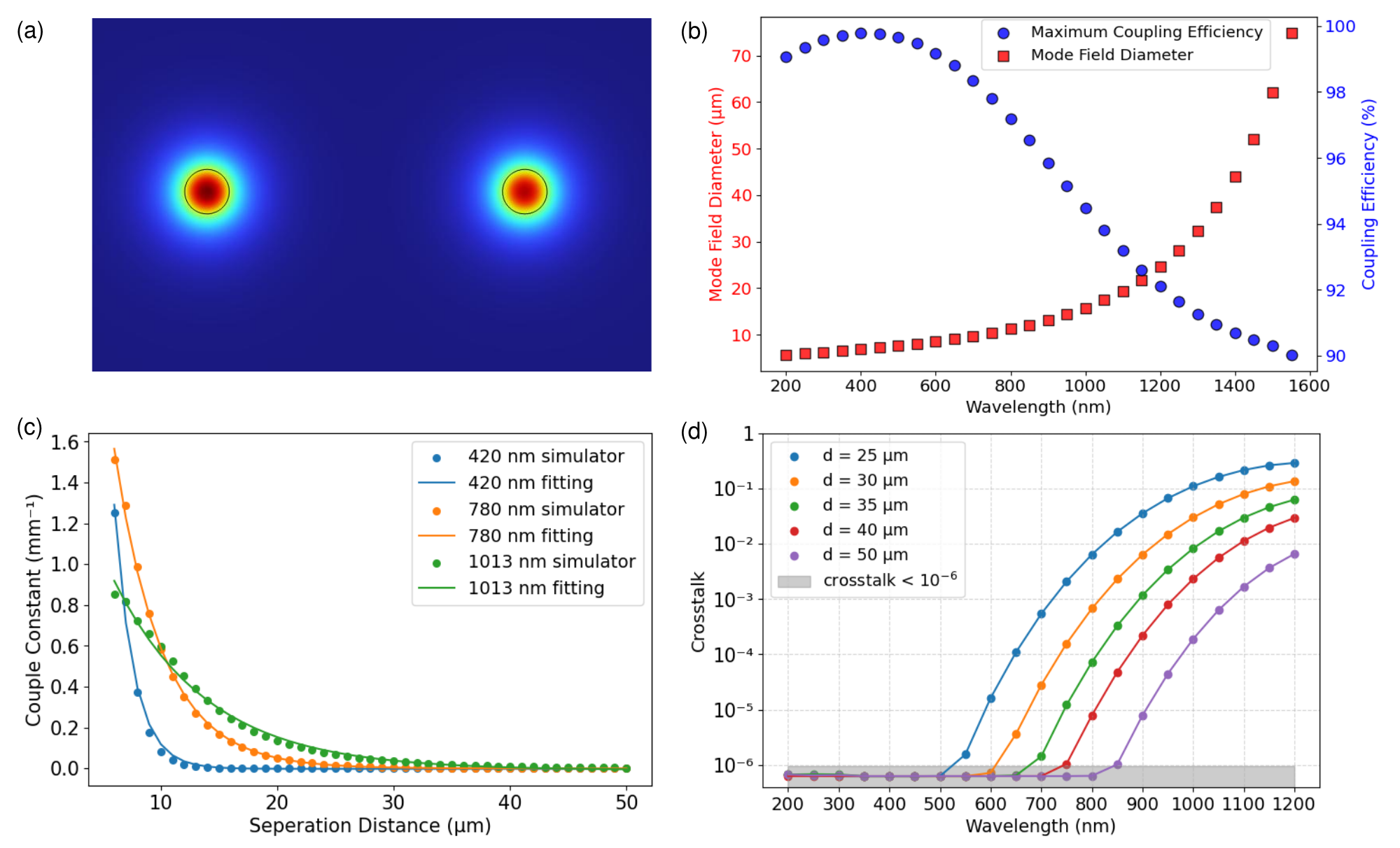}
\caption{Calculation of waveguides in photonic chip. 
(a) Mode field distribution of the fundamental waveguide mode. The black circle indicates stepwise refractive index boundaries. 
(b) Square points (left axis) are the wavelength-dependent mode field diameter (MFD) in the waveguide, while circular points (right axis) show the coupling efficiency between the waveguide and a Gaussian beam with the corresponding MFD. 
(c) Data points show the couple constant versus waveguide distance between two parallel waveguides at three specific wavelengths. Solid curves illustrate fitting results consistent with coupled-mode theory predictions. 
(d) Crosstalk characterization between two waveguides at a fixed effective coupling length of 5 mm. Colored markers quantify crosstalk across varying wavelengths with several coupling distances.
}
\label{fig4}
\end{figure*} 

\section{Implementation of Volcanic Architecture}
\subsection{3D Photonic Chip}

The OCM components can be realized through various approaches, including the spatial light modulator (SLM), DMD, fiber array, and 3D polymer waveguides. Here, we adapt a 3D photonic chip with waveguide structures fabricated by femtosecond laser writing techniques that modify the refractive index within glass substrates~\cite{Davis:96,Hirao1997,Gattass2008,Yang2024,Wang2024}.The precisely designed and fabricated three-dimensional waveguide structures enable arbitrary mapping between the input and output facet~\cite{Jovanovic2012,Marshall2009}. Compared with SLM, DMD, and fiber array, 3D photonic chip offers the potential of independent controlling on individual channels and the compatibility with planar photonic chips. 

To demonstrate the capability of the 3D photonic chip as an OCM, we fabricated a photonic chip with 49 laser direct writing waveguides, as shown in Fig.~\ref{fig3}(a). The chip architecture, detailed in Figs.~\ref{fig3}(b)-(d), features 3D pipeline waveguides that can be designed and fabricated at will, enabling arbitrary mapping to any given 2D qubit arrangement. For instance, the waveguides are arranged in a $1 \times 49$ array at the input facet for compatibility with standard commercial optical devices (such as the 1D waveguide array shown in Fig.~\ref{fig3}(a)), while forming a customized square lattice array, which fits the surface code arrangement~\cite{Fowler2012}, at the output facet. 

Figure~\ref{fig3}(e) illustrates how this photonic chip enables Volcano architecture implementation for both C-links and Q-links. For the C-link, we combine an AOD with the photonic chip for RF frequency-multiplexed addressing. When driven by multiple RF signals with equal frequency spacing, the AOD generates a corresponding beam array with deflection angles determined by the RF frequencies. The beam spacing precisely matches the 1D input ports of the OCM, allowing each beam to be routed to its designated qubit site through the 3D waveguide chip. This configuration establishes direct RF frequency-to-qubit site mapping: selecting RF frequencies determines which qubits are addressed, while modulating the amplitude and phase of each RF component implements the desired quantum operations. It is worth noting that the AOD allows fast switching time with a microsecond-scale rising edge, and also parallel control of hundred qubits by a single AOM. 

For the Q-link, the system operates in reverse to enable efficient photon collection. A high-NA objective lens collects fluorescence emitted from individual qubits to the chip's 2D output facet. As shown in Fig.~\ref{fig3}(a), the collected photons are routed to the chip's 1D input facet and further guided into a standard fiber array through a direct facet-to-facet coupling. Each fiber in the array collects photons emitted by its corresponding qubit, establishing a one-to-one mapping between qubit and fiber channels.  Such a Q-link enables parallel, independent detection or transmission of quantum signals from the qubit array. Notably, coupling between the photonic chip and fiber array requires only passive alignment, thus minimizing insertion loss for efficient quantum channel.

\subsection{Key Advantages of 3D Photonic Chips}

Other than the unique ability to engineer light propagation in three dimensions with sub-$\mathrm{\mu m}$ precision, the 3D photonic chip-based OCM holds many advantages when realizing the Volcano architecture for scalable QPUs. In the following, we examine five key advantages: 
\begin{enumerate}
    \item Massive scalability. The maximum number of waveguides in 3D photonic chips is primarily constrained by two critical factors: the minimum achievable waveguide spacing and the fabrication writing depth limit of the 2D facet. For example, a design with $\qty{50}{\mu m}$ waveguide spacing over a $1\,\mathrm{cm}$ chip length allows 200 columns of waveguides to be fabricated. Combined with a $1\,\mathrm{mm}$ depth capacity enabled by high-NA objective lenses~\cite{Huang2016}, the architecture supports 20 vertically stacked layers of waveguide. Through this scheme, the 3D photonic chip achieves a theoretical maximum of 4,000 independent channels, matching the state-of-the-art achievable qubit number with neutral atoms~\cite{Manetsch2024,luchaoyang} and trapped ions~\cite{Guo2024}.
        \item Gaussian-like mode profile. Figure~\ref{fig4}(a) displays the mode field distribution of the waveguide array, with circular waveguide cross-sections and a stepwise refractive index distribution ~\cite{Ams2005,Nolte2003}. Numerical results show that the mode profile closely matches a Gaussian beam over a wide range of wavelengths, except for significant deviation of beam profile the Gaussian-like distribution at long wavelengths, as shown in Fig.~\ref{fig4}(b). This degradation is caused by optical energy spreading out of the high-index waveguide region. This property enables efficient conversion to free-space Gaussian beams for coherent optical photon-qubit interactions and high-efficient fluorescence collection. Furthermore, the Gaussian-like mode profile allows direct fiber coupling with minimal loss, facilitating modular extensions via fiber connections. 
    \item   Ultralow loss. The propagation loss of laser-direct-written waveguides typically measures $0.055\,\mathrm{dB/cm}$ for 700\,nm~\cite{Colliard22}, corresponding to a photon loss rate of less than $2\%$ for a centimeter-scale chip. The waveguides maintain low loss over a broad spectral band from 300\,nm to 1550\,nm, encompassing the operating wavelengths of all major qubit platforms. 
    \item Low crosstalk. The application of this 3D photonic chip in OCM requires minimal inter-waveguide coupling to ensure low crosstalk between channels. Figures~\ref{fig4}(c) and (d) show the numerical results for the coupling between adjacent waveguides. For three selected wavelengths (420\,nm, 780\,nm, 1013\,nm), the couple constance between waveguides show a exponential decay $\kappa = \kappa_0 e^{-\frac{d}{d_0}}$ with respect to their separation $d$, where $d_0$ is the characteristic decay distance of the evanescent field. Using coupled-mode theory, the crosstalk for different wavelengths is calculated under an effective coupling length of $L = 5\,\mathrm{mm}$, defined as the physical length over which adjacent waveguides maintain a fixed parallel separation and interact via evanescent fields. For $d=30\,\mathrm{\mu m}$, crosstalk is smaller than $10^{-3}$ at visible wavelengths, while the crosstalk could be further suppressed when $d=50\,\mathrm{\mu m}$.   
    \item Compatibility with planar chips. The 1D-to-2D OCM can also be applied to bridges the gap between planar photonic integrated circuits with qubit array. Planar photonic chip features integrated modulator~\cite{Li2023,Zhang:25}, frequency converter~\cite{PhysRevLett.68.2153,PRXQuantum.5.010101}, and other functional devices~\cite{Li2022}, would further enhance the scalability of the Volcano architecture. 
\end{enumerate}

\begin{figure}[!t]
\centering
\includegraphics[width=\linewidth]{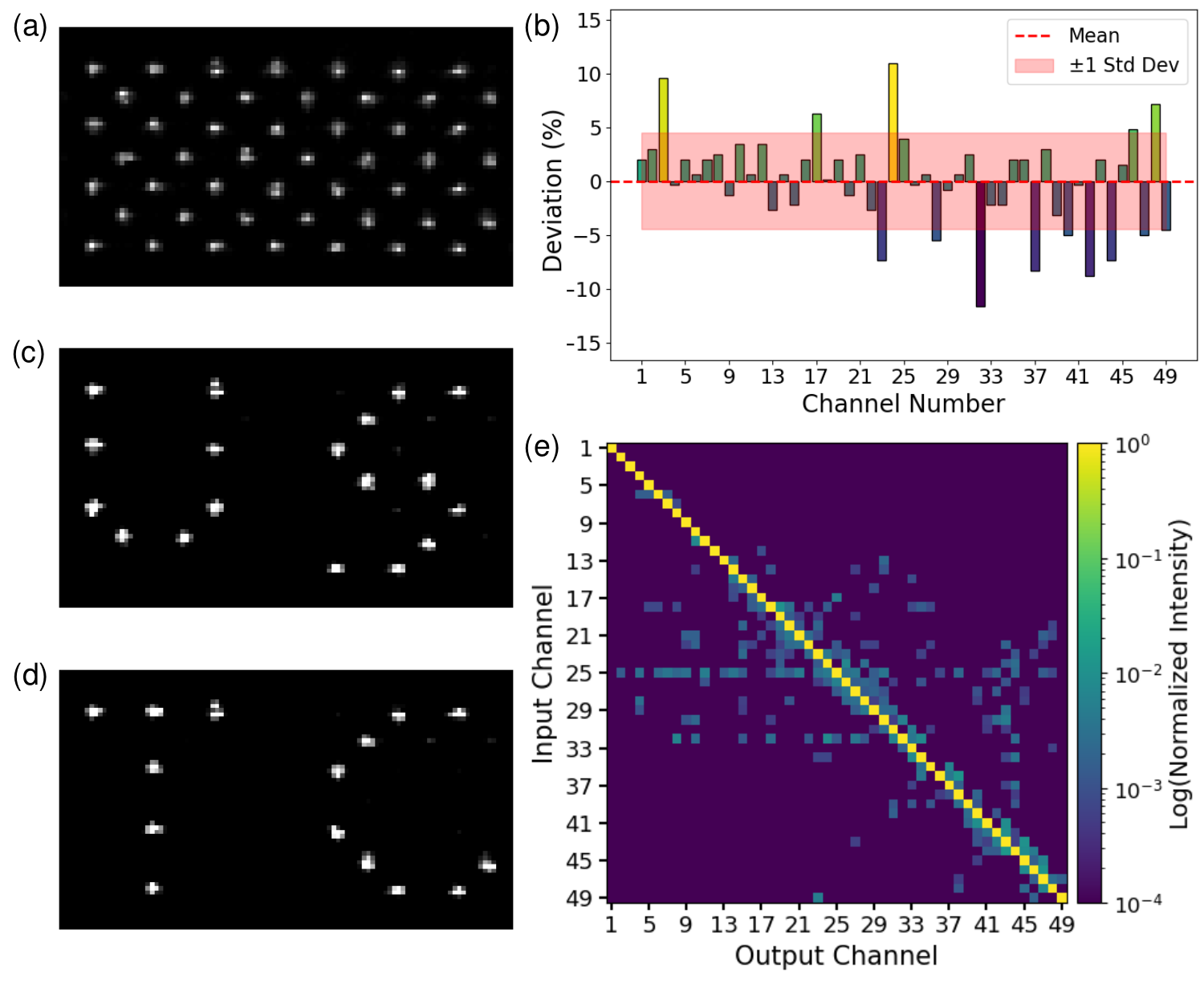}
\caption{
Characterization of waveguides in photonic chip. 
(a) Imaging results after 49 channels are fully activated, recorded by a CCD after guidance through the 3D photonic chip.
(b) Characterization of uniformity of 49 channels: the relative standard deviation (RSD) is $95.5\%$.
(c) Imaging result of the addressing demonstration “
US”.
(d) Imaging result of the addressing demonstration “TC”.
(e) Crosstalk matrix for the 49 channels, where rows indicate enabled channel IDs and columns represent the corresponding imaged channel IDs. Matrix entries quantify optical intensity measured in target columns when each row channels are individually enabled. The maximum, average nearest-input, and average nearest-output crosstalk values are measured as $1.198\%$ , $0.232\%$, and $0.158\%$, respectively. 
}
\label{fig5}
\end{figure}

\subsection{Proof-of-principle Demonstration}

Having established the theoretical advantages of 3D photonic chips, we now turn to experimental validation. We characterized the processed photonic chip using a setup analogous to Fig.~\ref{fig3}(e). In this configuration, an arbitrary waveform generator (QC-100, BZ-TEK) delivers up to 49 RF components to an AOD, which generates a 1D array of 49 optical beams at 420\,nm wavelength to the chip, and the output 2D addressing beam array pattern is captured by a CCD. This configuration tests the complete C-link: electrical signals$\rightarrow$RF modulation on AOD$\rightarrow$1D optical beams$\rightarrow$3D chip$\rightarrow$2D qubit array addressing. 

Figure~\ref{fig5}(a) presents the result when simultaneously activating all 49 channels for parallel addressing of all sites. While the beam profiles at most sites exhibit Gaussian-like spots manifests single-mode operation, some sites exhibit irregular patterns, indicating the excitation of higher-order modes in the waveguide chip due to fabrication imperfections or mis-alignment of input beams to the chip. To mitigate the inter-modulation in the AOD, a fixed frequency interval was maintained between adjacent RF components. Systematic scanning of both frequency intervals and individual RF amplitudes enabled optimization, achieving an optical spot array with enhanced uniformity. Fig.~\ref{fig5}(b) illustrates the uniformity between channels, where the light intensity across all sites yields a relative standard deviation (RSD) of 95.5\%, demonstrating consistent intensity profiles throughout the array. Selective addressing of qubit arrays is achieved by dynamically adjusting RF component parameters between the AWG and AOD. Figures~\ref{fig5}(c) and (d) demonstrate the arbitrary addressing capability through optical spot arrays forming ``US" and ``TC" patterns, showcasing the programmable control of the OCM for the C-link.  

The characterization of crosstalk between waveguides is tested by measuring the complete $49\times49$ crosstalk matrix, as shown in Fig.~\ref{fig5}(e). The results are obtained by sequentially activating each channel while recording the optical power at all 49 output sites. The maximum crosstalk is $1.198\%$, the average nearest-neighbor crosstalk for the sites at 1D input facet is $0.232\%$, and the average nearest-neighbor crosstalk for the sites at 2D output facet is $0.158\%$. The measured difference between average nearest neighbor crosstalk at the input and output facets arises from beam-profile mismatch between the AOD input beam spot and the waveguide mode or misalignment between them. The average non-nearest-neighbor crosstalk is $0.020\%$, primarily attributed to leakage in the chip substrate. Additionally, we characterize the insertion loss through fiber array coupling to the photonic chip, demonstrating a total loss of $2.5\,\mathrm{dB}$ which is acceptable for photon collection in Q-link applications. These imperfections could be mitigated through improving fabrication of the 3D waveguide chip and better optical design for mode matching~~\cite{Sotirova2024}.

\begin{figure}[!t]
\centering
\includegraphics[width=\linewidth]{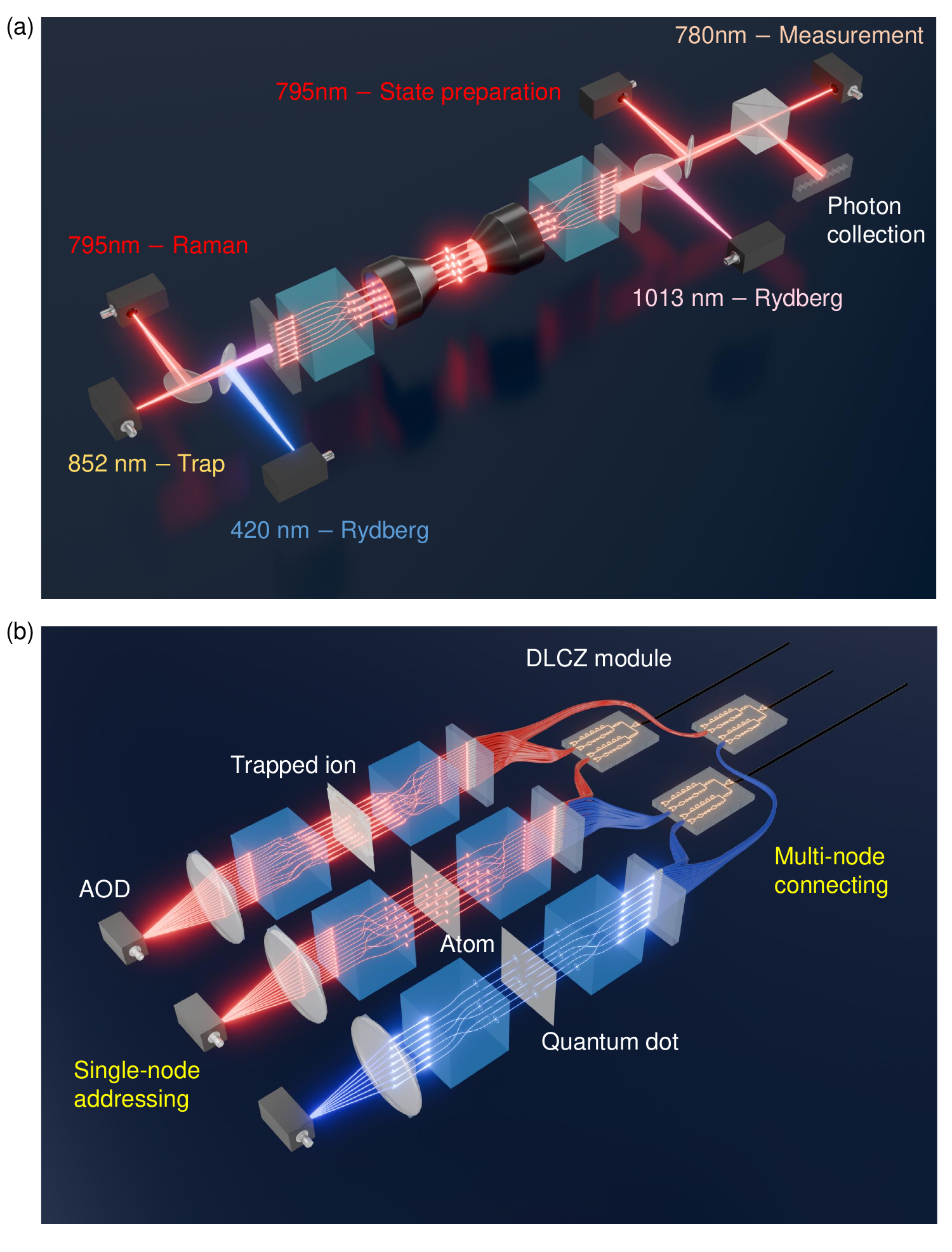}
\caption{\textbf{Implementation of the Volcano architecture in quantum computing and networking applications.} (a) Schematic of a neutral atom quantum processor utilizing the Volcano architecture. An optical tweezer array generated by the photonic chip traps individual atoms, allowing for arbitrary 2D qubit arrangements. The same photonic chip guides the addressing lasers for state preparation, single-qubit gates (Rx and Rz rotations), and two-qubit Rydberg gates, while also collecting the atomic fluorescence for parallel readout. (b) Elementary quantum network link enabled by the Volcano architecture. An ordered atomic array produces entangled photon pairs via collective light-matter interaction and photon collection by the photonic chip. After frequency conversion and transmission through optical fibers, the photonic entanglement can be mapped onto remote quantum nodes consisting of trapped ions and quantum dots, establishing a heterogeneous quantum network. The inset shows the energy level diagrams for neutral atom (Rb-87), trapped ion (Yb-171), and quantum dot systems, illustrating the diverse wavelengths involved. }
\label{fig6}
\end{figure} 

\section{Discussions}

The fundamental limits of any OCM technology arise from three physical constraints: the diffraction limit (setting minimum channel spacing), material absorption (defining operational wavelength range), and fabrication precision (determining channel uniformity and crosstalk). Our 3D photonic chip approach optimally balances these constraints, achieving thousands of channels through three-dimensional routing while maintaining the low loss and broadband operation essential for quantum applications. Beyond our demonstrated approach, alternative OCM technologies offer complementary capabilities worth exploring. Fiber array bundles provide perfect mode matching and mechanical flexibility, though manual assembly limits them to specific geometric patterns and moderate channel counts~\cite{Li2024}. SLMs and DMDs enable dynamic reconfiguration by real-time adjustment of the mapping but suffer from inherent crosstalk when mapping between dimensions~\cite{Zhang2024}. Emerging metasurface technology promises extreme miniaturization through subwavelength engineering, potentially integrating OCM functionality directly onto quantum chips, though current fabrication challenges limit uniformity and efficiency. The future likely lies in hybrid approaches: 3D waveguides with integrated active elements for dynamic routing, or hierarchical OCM architectures where multiple stages provide both coarse and fine addressing. These advances could push beyond static mapping to truly programmable optical interconnects.

The Volcano architecture is ready for immediate deployment in neutral atom quantum processors, with natural extensions to chip-to-chip interconnects and cross-platform quantum networks. Figure~\ref{fig6}(a) illustrates the neutral atom quantum computing system based on the Volcano architecture. The implementation leverages the 3D chip's broadband capability to route multiple optical functions through shared waveguide chips: 852\,nm light creates the trap array, 780\,nm enables fluorescence readout, 795\,nm drives state preparation and Raman transitions, while 420\,nm and 1013\,nm address different Rydberg states for two-qubit operations. This five-wavelength operation through a single chip design exemplifies the power of 3D waveguide chips. Specially, C-link enables independent and parallel switching and modulation of optical channels for precise control of individual atoms, which enables addressable light-shift to realize Z-axis rotation, Raman transition for X-axis rotation, Rydberg transition for two-qubit gates, with the parallel gate implementation and $\mathrm{\mu s}$-level switching time enabled by the RF-to-qubit multiplexing control. The Q-link enables efficient interface between array of SPDs and qubits, thus allowing high‑fidelity addressable readout of qubits states at around $\qty{100}{\mu s}$~\cite{k2w2-83kc}, which is essential for mid-circuit measurement and feedforward control for dynamics quantum circuits and quantum error correction. 

The ultimate vision enabled by the Volcano architecture is a full-stack quantum network where modular atomic quantum processors with tens to hundreds of qubits act as multi-functional nodes that can generate and store entanglement in quantum memories. Figure~\ref{fig6}(b) shows the potential of the modular design of Volcano architecture for further extension. The entangled atoms in the ordered atomic arrays can generate entangled photon pairs, with the Volcano chip efficiently collecting these photons into fiber channels. After frequency conversion to telecom wavelengths, this quantum light can be distributed to remote nodes containing different qubit types, including neutral atoms of different species, trapped ions, and quantum dots, each interfaced through their own Volcano chips. Alternatively, the entanglement between qubits in the same array over long distances or qubits separated in different nodes can be entangled by the DLCZ-type scheme. This creates a ``quantum internet” for potential quantum communication, distributed quantum computing, and quantum sensing applications.

\section{Conclusion}
We have presented the Volcano architecture for scalable quantum processing units (QPUs) and identified the fundamental challenges in realizing a scalable classical link (C-link) for qubit control and quantum link (Q-link) for photon collection. To overcome this scaling bottleneck, we introduce optical channel mapping (OCM) devices that realize arbitrary mapping between two-dimensional qubit arrays and one-dimensional optical channels from commercial components. As a proof-of-principle demonstration, we implement a 49-channel OCM by a three-dimensional photonic chip, which achieves a nearest neighbor average crosstalk smaller than $0.3\%$ and a uniformity of $95.5\%$, validating its feasibility for C-link and Q-link implementation. The OCM can operate across a broad wavelength (300\,nm - 1100\,nm), enabling crucial capabilities for scaling up QPUs: parallel addressing of arbitrary qubit array configurations, addressable single-shot readout, and rapid reconfiguration of control sites. It can be further extended to the network level by providing an efficient optical interface for entanglement distribution and quantum state transfer between nodes~\cite{ZhangYL2024}, thus making it compatible with modular quantum processor architectures. Therefore, the Volcano architecture establishes a practical pathway toward large-scale quantum computing and quantum network.

\section*{Acknowledgements}
\begin{acknowledgments}
We would like to thank Hangzhou Biaozhang Electronic Technology Co., Ltd. for their assistance in setting up electronics equipment. This work was funded by the National Key R\&D Program (Grant No.~2021YFA1402004), the National Natural Science Foundation of China (Grants No. U21A20433, U21A6006, 92265210, and 92265108), and Innovation Program for Quantum Science and Technology (Grant No.~2021ZD0300203), and the Natural Science Foundation of Anhui Province (Grant No. 2408085QA017). This work was also supported by the Fundamental Research Funds for the Central Universities and USTC Research Funds of the Double First-Class Initiative. The numerical calculations in this paper have been done on the supercomputing system in the Supercomputing Center of University of Science and Technology of China. This work was partially carried out at the USTC Center for Micro and Nanoscale Research and Fabrication.
\end{acknowledgments}

\vspace{\baselineskip} 
\noindent\textbf{Conflict of Interest}\\
The authors declare that they have no conflict of interest.

\newpage

\clearpage

\end{document}